\documentclass[prl,aps,twocolumn,amsfonts]{revtex4}

\usepackage{graphicx}
\usepackage{epsfig}

\def\qed{\leavevmode\unskip\penalty9999 \hbox{}\nobreak\hfill
     \quad\hbox{\leavevmode  \hbox to.77778em{%
               \hfil\vrule   \vbox to.675em%
               {\hrule width.6em\vfil\hrule}\vrule\hfil}}
     \par\vskip3pt}

\def\ibb #1{\leavevmode\hbox{\kern.3em\vrule
     height 1.5ex depth -.1ex width .4pt\kern-.3em\rm#1}}

\newcommand{\ba}[2]{\left(\begin{array}{#1}#2\end{array}\right)}
\newcommand{\tr}[1]{{\rm Tr}\left(#1\right)}
\newtheorem{theorem}{Theorem}

\begin{document}

\title{Optimal teleportation with a mixed state of two qubits.}

\author{Frank Verstraete$^{ab}$ and Henri Verschelde$^b$\\
 $^a$Department of Electrical Engineering (ESAT-SISTA), KULeuven,
Belgium\\
 $^b$Department of Mathematical Physics and Astronomy, Ghent University,
 Belgium}
\date{October 30, 2002}

\begin{abstract}
We consider a single copy of a mixed state of two qubits and
derive the optimal trace-preserving local operations assisted by
classical communication (LOCC) such as to maximize the fidelity of
teleportation that can be achieved with this state. These optimal
local operations turn out to be implementable by one-way
communication, and always yields a teleportation fidelity larger
than $2/3$ if the original state is entangled. This maximal
achievable fidelity is an entanglement measure and turns out to
quantify the minimal amount of mixing required to destroy the
entanglement in a quantum state.
\end{abstract}

 \maketitle
The basic resource in quantum information
theory consists of maximally entangled qubits or Bell states. This
stems from the fact that the additional resource of entanglement
enables to implement all possible global quantum operations
locally by making use of the concept of quantum teleportation
\cite{BBC93}. Perfect teleportation is only possible when
maximally entangled states are available. In practical situations
however, we have to deal with mixed states due to the undesired
coupling of the quantum states with the environment. In this
letter we address the following basic question: given a mixed
state of two qubits, what is the maximal teleportation fidelity
that can be obtained with this state allowing all possible
trace-preserving local operations assisted by classical
communication (LOCC) or all possible filtering operations (SLOCC)?
We give a complete answer to those questions.

If only local unitary operations are allowed, then the Horodecki's
\cite{HHH99} proved that the optimal teleportation fidelity $f$ is
a linear function $f=(2F+1)/3$ of the maximal singlet fraction or
fidelity $F$ \cite{BDS96}, which is defined as the maximal overlap
of a state with a maximally entangled state:
\[F(\rho)=\max_{|\psi\rangle={\rm
ME}}\langle\psi|\rho|\psi\rangle.\] Therefore the problem of
maximizing the teleportation fidelity is equivalent to maximizing
the maximal singlet fraction of a mixed state of two qubits. This
problem is also of great interest in  the context of distillation
protocols \cite{BBP96b,BDS96,DEJ96,DVV02}, as the distillation of
barely entangled states occurs through the use of recurrence
schemes which gradually enhance the fidelity under the condition
that the initial fidelity $F$ exceeds $1/2$. The maximal fidelity
for separable states is indeed given by $F=1/2$.  Surprisingly,
there exist entangled states whose fidelity is lower than this
value \cite{VV02b}, but the Horodecki's \cite{HHH97} proved that
local filtering operations can always be chosen such that, with a
finite probability, a state with fidelity exceeding $1/2$ is
obtained if the original state was entangled. In this letter we
will prove the stronger result that  a mixed  state of two qubits
is entangled if and only if there exist trace-preserving LOCC
operations such that the fidelity $F^*$ of the LOCC-processed
state exceeds $1/2$. This answers the following question raised by
Badziag and Horodecki \cite{BHH00} in the case of two qubits: "Can
any entangled state provide better than classical fidelity of
teleportation?"

The present work also sheds new light on the open problem of
characterizing the class of local operations that can physically
be implemented on a system.  Finding a parametrization of the
class of LOCC operations turns out to be very difficult. The class
of PPT-operations \cite{Rai01} however, related to the concept of
partial transposition \cite{Per96,HHH96}, is very easy to
characterize, but contains operations that cannot be implemented
locally. In this letter, we will consider the problem of
maximizing the fidelity under the action of all PPT-operations.
Surprisingly, it will turn out that the optimal PPT-protocol is
always physically implementable: this is supporting evidence for
the fact that the class of PPT-operations yields a good
approximation of the class of LOCC operations.

In a first part, the optimal local filtering operations
\cite{Gi96} such as to maximize the fidelity of the filtered state
are derived. The optimal filter is the one that transforms the
state into a unique Bell diagonal form. This provides a
quantitative way of characterizing the qualitative result of the
work of Horodecki \cite{HHH97}. The drawback of filtering
operations is the fact that these operations can only be
implemented with a certain probability. It is therefore an
interesting question whether trace preserving local operations can
also enhance the fidelity. In a surprising paper of Badziag et al.
\cite{BHH00}, it was shown that there exist mixed states with
fidelity smaller than $1/2$, for which local trace-preserving
protocols exist that transform this state into a state with
fidelity larger than $1/2$ without the help of classical
communication. Motivated by this example, we looked for the
optimal LOCC protocols such as to transform an entangled state
into one with fidelity as large as possible allowing classical
communication. We prove that the optimal trace-preserving protocol
for maximizing the fidelity of a given state always belongs to a
very simple class of 1-LOCC operations, and provide a constructive
way of obtaining this optimal (state-dependent) LOCC operation. We
conclude by giving a geometrical interpretation of the maximum
achievable fidelity by LOCC, revealing an interesting connection
with the robustness of entanglement \cite{VT99}.

Let us now state the first theorem of this letter:
\begin{theorem}
The bipartite local filtering operations probabilistically
bringing an entangled mixed state of two qubits to a state with
the highest possible fidelity are given by the filtering
operations bringing the state into its unique  Bell-diagonal
normal form\cite{VDD01b}, yielding a fidelity larger than $1/2$.
\label{thof}
\end{theorem}
{\em Proof:} In \cite{VDD01b} it was proven that the local
filtering operations maximizing the concurrence \cite{Woo98} of a
state are given by the local operations bringing the state into
its unique Bell diagonal normal form, and that a state is
entangled if and only if its normal form is entangled. The
fidelity of a state is bounded above by $F(\rho)\leq
(1+C(\rho))/2$ \cite{VV02b} with $C(\rho)$ the concurrence, and
for Bell diagonal states the equality holds. It is moreover
trivial to check that the fidelity of an entangled Bell diagonal
state exceeds $1/2$, which ends the proof.\qed

Next we want to investigate if there always exist trace-preserving
local operations such that the fidelity of the obtained state
exceeds $1/2$ if the original state is entangled. The crucial
point is to incorporate the previously described filtering
operation as part of a trace preserving LOCC operation. The idea
is that it is always possible to make a trace-preserving LOCC
operation out of a SLOCC filtering operation by making a pure
separable state if the state did not pass the filter. Then with a
certain probability a Bell diagonal state $\rho_f$ arises with
fidelity exceeding $1/2$, and with the complementary probability a
pure separable state $|\chi\rangle$ can be created having fidelity
equal to $1/2$ (note that $|\chi\rangle$ must be chosen such that
$|\langle \chi|\psi\rangle|^2=1/2$ with $|\psi\rangle$ the
maximally entangled state obeying $F(\rho_f)=\langle
\psi|\rho_f|\psi\rangle$). This proves that for each entangled
mixed state of two qubits there exists a trace-preserving 1-LOCC
protocol that transforms it into a state with fidelity larger than
$1/2$.

Let us now try to optimize the trace-preserving operation used in
the protocol just described such as to maximize the fidelity of a
given state. Note that in general the optimal filter of theorem
\ref{thof} will not be optimal in the trace-preserving setting as
in that case the probability of obtaining the state was not taken
into account. The setting is now as follows: we want to find the
filter, such that the probability of success $p_{AB}$ of the
filter multiplied by the fidelity $F$ of the state coming out of
this filter, plus $(1-p_{AB})$ times the fidelity of the pure
separable state given by $1/2$, is maximal. For given filter
$-I\leq A,B\leq I$, the cost-function $K_{AB}$ is therefore given
by
\[K_{AB}=p_{AB}F(\rho_f)+\frac{1-p_{AB}}{2}\] where
\begin{eqnarray*}p_{AB}&=&\tr{(A\otimes B)\rho(A\otimes B)^\dagger}\\
\rho_f&=&\frac{(A\otimes B)\rho(A\otimes B)^\dagger}{p_{AB}}
\end{eqnarray*}
Now some tricks will be applied. Due to the presence of $A,B$, we
can replace $F(\rho_f)$ by $\langle \psi|\rho_f|\psi\rangle$ with
$|\psi\rangle=(|00\rangle+|11\rangle)/\sqrt{2}$, and we use the
fact that the trace of the product of two matrices is equal to the
trace of the product of the partial transpose of two matrices.
This leads to the following expression (see also the proof of
theorem 1 in \cite{VV02b}):
\begin{equation}
K_{AB}=\frac{1}{2}-\langle\psi|(C\otimes
I)\rho^{\Gamma}(C^\dagger\otimes
I)|\psi\rangle\label{eqlb}\end{equation} where
$C=B^\dagger\sigma_y A$ and $\rho^{\Gamma}$ is the shortcut
notation for the partial transpose with respect to the system $B$.
This cost-function has to be maximized over all complex $2\times
2$ matrices $-I\leq A,B\leq I$, and this leads to a lower bound on
the maximum achievable fidelity by LOCC operations. Note that the
considered operations can always be implemented using one-way
communication (1-LOCC), as one can always choose $B=I$ without
loss of fidelity.

An upper bound can be obtained by using the techniques developed
by Rains \cite{Rai01}. Indeed, if we enlarge the class of allowed
operations from  trace-preserving LOCC operations to
trace-preserving PPT-operations, a simple optimization problem
arises. A quantum operation $\Lambda$ is PPT if and only if the
dual state $\rho_\Lambda$ associated to this operation
\cite{Jam72,HHH99,CDK01,Rai01,VV02a} is PPT. The dual state
$\rho_\Lambda$ corresponding to a map $\Lambda$ on two qubits is
defined in a $2\otimes 2\otimes 2\otimes 2$ Hilbert space and the
following relation holds:
\[\left(\Lambda(\rho)\right)^T_{A'B'}=4{\rm
Tr}_{AB}\left(\rho_\Lambda^{AA'BB'}(\rho_{AB}\otimes
I_{A'B'})\right).\] An upper bound on $F^*$ can now be obtained by
considering the following optimization problem: maximize
\[4\tr{\rho_\Lambda(\rho\otimes
|\psi\rangle\langle \psi|)}\] under the constraints
\begin{eqnarray*}
\rho_\Lambda&\geq &0\\
\rho_\Lambda^{T_{BB'}}&\geq &0\\
4{\rm Tr}_{A'B'}(\rho_\Lambda)&=&I_{AB}
\end{eqnarray*} and  with $|\psi\rangle$ a maximally
entangled state. Here the notation $\rho^{T_{BB'}}$ denotes the
partial transpose with respect to the systems $B$ and $B'$. This
is a semidefinite program and can easily be solved numerically
with guaranteed convergence \cite{VB96}. Exploiting symmetries
however, it is possible to reduce the complexity drastically.
Indeed, $|\psi\rangle$ remains invariant under a twirl operation
and this twirl can be applied to $\rho_\Lambda$, leading to a
state of the form:
\[\rho_\Lambda=\frac{1}{16}\left(I_4\otimes I_4+(4X-I_4)\otimes
(4|\psi\rangle\langle\psi|-I_4)\right).\] Here $X$, a $4\times 4$
matrix, is subject to convex constraints $I/6\leq X\leq I/2$ and
$0\leq X^\Gamma\leq I/3$. Doing the substitution $X\rightarrow
(I-X^\Gamma)/3$, this optimization problem reduces to the
following semidefinite program (see also Rains \cite{Rai01}):
maximize
\begin{equation}1/2-\tr{X\rho^\Gamma}\label{equb}\end{equation} under the constraints
\begin{eqnarray*}
0&\leq&X\leq I_4\\
-\frac{I_4}{2}&\leq & X^\Gamma\leq \frac{I_4}{2}.
\end{eqnarray*}
Note that the constraint $-\frac{I_4}{2}\leq X^\Gamma$ will
automatically be satisfied if the other three constraints are
satisfied: this follows from the fact that $X^\Gamma$ has at most
one negative eigenvalue $\lambda_-$ and that $|\lambda_-|\leq
\max(\lambda(X^\Gamma))$ \cite{VAD01b}. Suppose now that $X$
fulfills the constraints and has rank larger than one. Then $X$
has a separable state $S$ in its support, as each two-dimensional
subspace contains at least one separable state \cite{LS98}.
Consider now $y^2$ the largest real positive scalar such that
$X-y^2S\geq 0$. It is easy to verify that the matrix $Y=X-y^2S$
also fulfils the four constraints, as $S^\Gamma$ is positive due
to its separability. Moreover the value
$\tr{S\rho^\Gamma}=\tr{S^\Gamma\rho}$ with $S$ separable and
$\rho$ entangled is assured to be positive. Therefore the matrix
$Y$ will yield a larger value of the cost-function. This argument
implies that the maximal value of the cost-function will be
obtained for $X$ rank one. $X$ can therefore be written in the
form:
\[X=(A\otimes I_2)|\psi\rangle\langle\psi|(A^\dagger\otimes I_2),\]
and the constraints become $-I_2\leq A\leq I_2$.

But then the variational characterization of the upper bound
(\ref{equb}) becomes exactly equal to the variational
characterization of the lower bound (\ref{eqlb})! This is very
surprising as it implies that the proposed 1-LOCC protocol used in
deriving the lower bound was actually optimal over all possible
LOCC protocols. We have therefore proven:
\begin{theorem}
The optimal trace-preserving LOCC protocol maximizing the fidelity
of a given state $\rho$ consists of a 1-LOCC protocol where one
party applies a state-dependent filter. In case of success, the
other party does nothing, and in case of failure, both parties
make a pure separable state. The optimal filter and fidelity $F^*$
can be found by solving the following convex semidefinite program:
maximize
\[F^*=\frac{1}{2}-\tr{X\rho^\Gamma}\]
under the constraints:
\begin{eqnarray*}
0&\leq&X\leq I_4\\
-\frac{I_4}{2}&\leq&X^\Gamma\leq\frac{I_4}{2}.
\end{eqnarray*}
$F^*>1/2$ if $\rho$ is entangled and the optimal $X_{opt}$ will be
of rank 1, and the filter $A$ can be obtained by making the
identification
\[X_{opt}=(A\otimes I_2)|\psi\rangle\langle\psi|(A^\dagger\otimes I_2)\]
with $|\psi\rangle=(|00\rangle+|11\rangle)/\sqrt{2}$.
\end{theorem}
This theorem gives us the optimal way of using a mixed state of
two qubits for teleportation: the maximal possible fidelity will
be obtained when the state is first subjected to the optimal LOCC
protocol.

The given semidefinite program can be solved exactly if $\rho$ has
some symmetry. Indeed, if $\rho^\Gamma$ remains invariant under
certain symmetry operations, the optimal $X$ can always be chosen
such that it has the same symmetry (this follows from a similar
argument as the one used during the twirling step in the proof).
As an example, we will calculate $F^*$ for the family of states
\begin{equation}\rho(F)=F|\psi\rangle\langle\psi|+(1-F)|01\rangle\langle 01|\label{sfdkfsd}\end{equation} with
$F\geq 1/3$ the fidelity of the state (these are precisely the
states with minimal fidelity for given concurrence \cite{VV02b}).
The symmetries under transposition and under the local operations
$\sigma_z\otimes\sigma_z$ and ${\rm diag}[1,i]\otimes{\rm
diag}[1,i]$ imply that $X$ will be real and of the form
\[X=\ba{cccc}{x_1&0&0&0\\0&x_2&x_3&0\\0&x_3&x_4&0\\0&0&0&x_5}.\]
Moreover $x_1$ and $x_5$ will be equal to zero in the case of an
optimal $X$ as otherwise $X$ cannot be rank $1$, and a simple
optimization problem remains. The optimal filter is readily
obtained as $A={\rm diag}[F/(2(1-F));1]$, and the maximal
achievable fidelity $F^*$ becomes equal to:
\begin{eqnarray*}
F^*(\rho(F))=&\frac{1}{2}\left(1+\frac{F^2}{4(1-F)}\right)
\hspace{.3cm}&({\rm if\hspace{.1cm}} 1/3\leq F\leq 2/3)\\
F^*(\rho(F))=&F\hspace{.3cm}&({\rm if\hspace{.1cm}} F\geq 2/3)
\end{eqnarray*}
Note that for $F\geq 2/3$, no LOCC protocol exists that can
increase the fidelity for this class of states. This is  true in
general: for high fidelities, the fidelity is  very close to
$(1+N)/2$ with $N$ the negativity \cite{VV02b} which is an
entanglement monotone \cite{Vid00} and can therefore not be
increased by LOCC operations.

\begin{figure}
\begin{center}
    \scalebox{.4}{\includegraphics{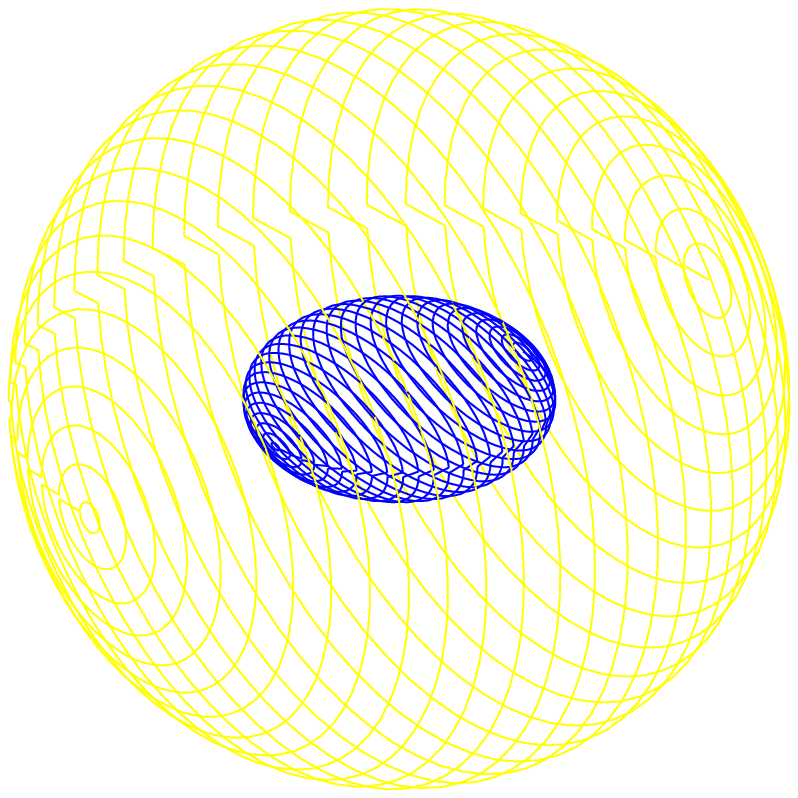}}
\scalebox{.4}{\includegraphics{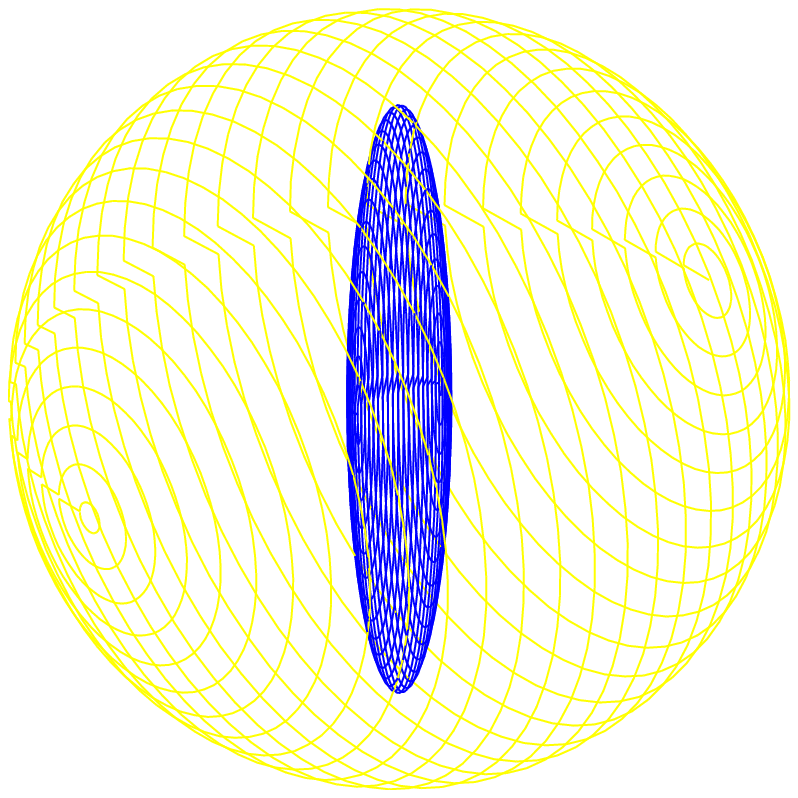}}
\scalebox{.4}{\includegraphics{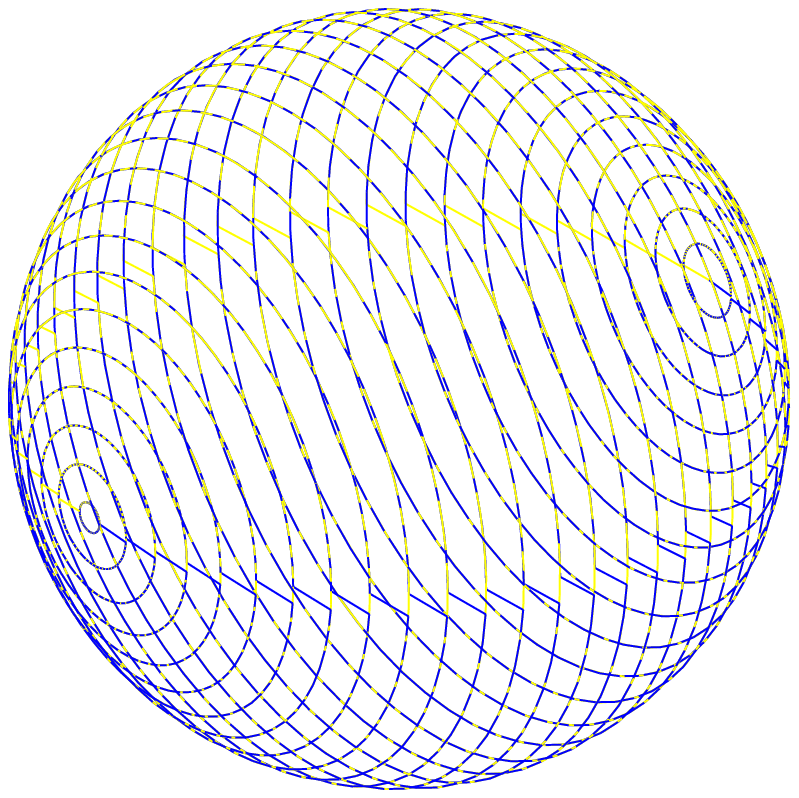}} \caption{The image of
the Bloch sphere induced by the teleportation channel with the
state $\rho(0.4)$ (eq. \ref{sfdkfsd}) under optimal LU (left),
LOCC (right) and SLOCC (middle) local preprocessing.}
    \label{figbstele}
\end{center}
\end{figure}

A quantum state used for teleportation is a special kind of unital
or bistochastic quantum channel (see e.g. \cite{HHH99,BB01}). A
unital quantum channel is completely characterized by looking at
the image of the Bloch sphere under the action of the
channel\cite{KR01}. In figure \ref{figbstele}, we depict the
images of the Bloch sphere under the action of the teleportation
channel obtained by the states $\rho(F)$ of eq.  \ref{sfdkfsd}
with $F=0.4$ when the  following preprocessing was done: 1.
optimal LU-preprocessing (implementation of the optimal local
unitaries such as to maximize singlet fraction \cite{HHH99}) ; 2.
optimal trace-preserving LOCC transformations (theorem 2); 3.
optimal filtering operations (theorem 1). This gives a nice
illustration of the obtained results.

For general states, no analytical method for obtaining an
expression of $F^*$ is known, and a (simple) semidefinite program
has to be solved. It is however easy to obtain an explicit lower
bound on the optimal $F^*$ in terms of the negativity and the
concurrence of the original state. This lower bound is obtained by
choosing $X$ to be a constant times the subspace spanned by the
negative eigenvector $v_-$ of $\rho^\Gamma$. This constant has to
be chosen such that the largest eigenvalue of $X^\Gamma$ does not
exceed $1/2$, and it can be shown that this implies that this
constant is equal to $1/(1+\sqrt{1-C^2_{v_-}})$ with $C_{v_-}$ the
concurrence of $v_-v_-^\dagger$. Using the variational
characterization of the concurrence of a mixed state
\cite{VDD01b,VDD02a}, it is furthermore easy to prove that
$C_{v_-}\geq N(\rho)/C(\rho)$. Putting all the pieces together, we
arrive at the following lower bound for the maximum achievable
fidelity $F^*$ for an arbitrary state $\rho$:
\[
\frac{1}{2}\left(1+\frac{N(\rho)}{1+\sqrt{1-\left(\frac{N(\rho)}{C(\rho)}\right)^2}}\right)\leq
F^*(\rho)\leq \frac{1}{2}(1+N(\rho)).\]

The upper bound follows from the fact that the fidelity is bounded
above by $(1+N)/2$ \cite{VV02b} which is an entanglement monotone
and can therefore not be increased by LOCC operations.

Before concluding, we will show that the maximum achievable
fidelity $F^*$ belongs to the class of entanglement measures
measuring the robustness of entanglement \cite{VT99,VW02}. To that
purpose, we use the fact that to each formulation of a
semidefinite program, there exists a dual formulation \cite{VB96}
that yields exactly the same value for the extremum. The dual of
(\ref{equb}) can be shown to reduce to: minimize
\[G=\frac{1}{2}+\frac{1}{2}\tr{Z}\]
subject to the constraints
\begin{eqnarray*}
Z&\geq &0\\
(\rho+Z)^\Gamma&\geq&0.\end{eqnarray*} Defining the state
$\rho_Z=Z/\tr{Z}$, this problem is equivalent to: minimize
\[G=\frac{1}{2(1-p)}\]
over all $0\leq p<1$ and over all states $\rho_Z$, subject to the
constraint that the state $\rho'$
\[\rho'=(1-p)\rho+p\rho_Z\]
is separable. The minimum value obtained is the maximum achievable
fidelity $F^*$. As $1/(1-p)$ is monotonously increasing over
$0\leq p<1$, this problem amounts to finding the state $\rho_Z$
such that the weight in the mixture of this state with the
original state $\rho$ is minimal, under the constraint that this
mixture is separable. The maximal achievable fidelity $F^*(\rho)$
is therefore a measure of the minimal amount of mixing required of
$\rho$ with another state such that a separable state is obtained.

In summary, we have shown that the fidelity or maximal singlet
fraction is not an entanglement monotone, but can be made one by
defining a new fidelity $F^*$ as the maximal achievable one by
trace-preserving LOCC operations. These optimal operations were
completely determined, and this maximal achievable fidelity $F^*$
quantifies the minimal amount of mixing required for a quantum
state to destroy its entanglement. The optimal achievable
teleportation fidelity is given by $f^*=(2F^*+1)/3$.


\bibliographystyle{unsrt}

\end{document}